\documentclass[pra,aps,twocolumn,showpacs]{revtex4-1}
\usepackage{eurosym}
\usepackage{amsmath,amssymb,graphicx}
\usepackage{epsfig}
\usepackage{textcomp}
\usepackage{color}
\usepackage{epstopdf}
\usepackage{hyperref}

\hypersetup{pdfpagemode=UseNone,colorlinks=true,linktoc=page,pdfborder={0 0 0},linkcolor=blue,citecolor=blue}

\begin{document}

\title{Quantum versus classical chirps in a Rydberg atom}
\author{Tsafrir Armon and Lazar Friedland}
\email{lazar@mail.huji.ac.il}

\begin{abstract}
The interplay between quantum-mechanical and classical evolutions in a
chirped driven Rydberg atom is discussed. It is shown that the system allows
two continuing resonant excitation mechanisms, i.e., a successive two-level
transitions (ladder climbing) and a continuing classical-like nonlinear phase
locking (autoresonance). The persistent $1:1$ and $2:1$ resonances between
the driving and the Keplerian frequencies are studied in detail and
characterized in terms of two dimensionless parameters $P_{1,2}$
representing the driving strength and the nonlinearity in the problem,
respectively. The quantum-mechanical rotating wave and the classical single
resonance approximations are used to describe the regimes of efficient
classical or quantum-mechanical excitation in this two-parameter space.
\end{abstract}

\maketitle

\affiliation{Racah Institute of Physics, Hebrew University of Jerusalem, Jerusalem 91904, Israel}

%%%%%%%%%%%%%%%%  INTRODUCTION   %%%%%%%%%%%%%%%%%%%%%%%%%%%%%

\section{Introduction}

\label{introduction}

Rydberg atoms possess many unique properties. With their large principal
quantum number, $n\gg 1$, they exhibit a long radiative lifetime (scaling as $n^{3}$%
), a large orbital radius and dipole moment (scaling as $n^{2}$) and more (see
\cite{Lit1} and references therein). As a result, they show great promise in
many applications including quantum nondemolition measurements of photons
\cite{Lit2}, digital communication \cite{Lit3}, measurement of microwave
fields \cite{Lit4,Lit5}, quantum information (see \cite{Lit6} for a
comprehensive review), and more. The ability to manipulate and control
Rydberg atoms is thus of great importance.

Of particular interest are circular Rydberg states (CRSs), i.e., Rydberg atoms
in states with $l=n-1$, where $l$ is the orbital quantum number and $m$, the magnetic quantum number, can take any value $\left|m\right|\le n-1$. Such states
have the longest radiative lifetime and magnetic moment \cite{Lit7}, which
makes them better suited for many applications. Various techniques for the
creation of CRSs have been proposed and implemented over the years \cite%
{Lit7, Lit8,Lit9,Lit10}. CRSs have been used in several innovative
advances in cavity quantum electrodynamics \cite{Lit11,Lit12} and are
proposed for future applications like quantum simulators of many-body
physics \cite{Lit13}.

In recent years, chirped frequency drives were studied as a tool for control
and manipulation of various oscillatory systems, including Rydberg atoms
\cite{Lit14,Gros,Lit14a,Lit15,Lit15a}. In many cases, the response of the
system to the chirped drive could take a quantum-mechanical, classical or
mixed form. In the classical limit, a persistent nonlinear phase locking
between the driver and the system, known as autoresonance (AR) \cite{Lit15b}%
, yields a continuing excitation. In contrast, in the quantum limit, the
system undergoes successive Landau-Zener (LZ) transitions \cite{Landau,Zener}%
, i.e., the quantum energy ladder climbing (LC). Both regimes of operation
have been demonstrated and used in atoms and molecules \cite%
{Lit16,Lit17,Lit18,Lit19,Lit19b}, Josephson junctions \cite{Lit20}, plasma
waves \cite{Lit21,Lit22}, discrete nonlinear systems \cite{DNLSE}, and cold
neutrons \cite{Lit23}.

In this work, we study the effects of a linearly polarized, chirped
frequency electric field on a Rydberg atom initialized in a CRS. Even though we usually associate large quantum numbers with the emergence of classical phenomena, we show that it is not a sufficient condition. Particularly, Rydberg atoms in a CRS, while having $n,l\gg 1$, can exhibit both classical and quantum-mechanical responses to the chirped drive.

We describe the characteristics of quantum-mechanical persistent resonance and compare it to the previously studied case of classical autoresonance \cite{Gros,1D}. With the use of a unified parametrization, the necessary conditions for each resonant regime are mapped, allowing one to easily determine what evolution should be expected for a given parameter choice, and when each regime is accessible.
The persistent $1:1$ and $2:1$ resonances between the
driving and the Keplerian frequency of the atom are discussed in detail.

The scope of the paper is as follows: In Sec. \ref{Sec1}, we introduce
the model and its parametrization. Section \ref{Sec2} characterizes the
resonant structure of the problem and Sec. \ref{Sec3} analyzes the
quantum and classical resonant regimes and the associated parameter space
for the $1:1$ resonance. Section \ref{Sec4} builds on Sec. \ref{Sec3} and
analyzes the $2:1$ resonance, while our conclusions are summarized in Sec. %
\ref{summary}.

%%%%%%%%%%%%%%%%  SEC I  %%%%%%%%%%%%%%%%%%%%%%%%%%%%%

\section{The Model \& Parameterization}

\label{Sec1}

We consider a Rydberg atom driven by an oscillating electric field of
constant amplitude and a down chirped frequency $\omega _{d}$ such that $%
d\omega _{d}/dt=-\alpha $, with $\alpha $ being a constant chirp rate. The
Hamiltonian of the problem $\hat{H}=\hat{H_{0}}+\hat{U}$ includes the usual
unperturbed part

\begin{equation}
\hat{H_{0}}=\frac{\hat{\vec{p}}^{2}}{2m_{e}}-\frac{e^{2}}{\hat{r}},
\label{H0}
\end{equation}%
and the driving part $\hat{U}=\varepsilon \cos {\phi _{d}}\hat{z}$, where $%
m_{e}$ and $e$ are the electron's mass and charge, $\varepsilon $ \ and $%
\phi _{d}=\int_{0}^{t} \omega _{d}\left(t'\right)dt'$ are the driving amplitude and phase
respectively, and the driving field is in the $z$ direction. The operator $%
\hat{z}$ is dimensionless, with the normalization constant included in $%
\varepsilon $. The eigenfunctions $\left\vert n,l,m\right\rangle $ of $\hat{%
H_{0}}$ satisfy $\hat{H_{0}}%
\left\vert n,l,m\right\rangle =E_{n}\left\vert n,l,m\right\rangle $ where $%
E_{n}=-R_{y}n^{-2}$ and $R_{y}$ is the Rydberg energy. Note that we
neglected the corrections to the energy due to other quantum defects \cite%
{Defect1,Defect2}, as they are fairly constant and do not have notable
consequences for this work. As an initial condition we consider a
single $\left\vert n_{0},n_{0}-1,m\right\rangle $ CRS.

The resonances in the problem are studied in detail in Sec. \ref{Sec2},
but their nature is important for the choice of a suitable parametrization.
These resonances correspond to a $q:1$ ratio between $\omega _{d}$ and the
Keplerian frequency (approximately given by $\frac{dE_{n}}{dn}/\hbar $), and
they affect transitions between the states $\left\vert n,l,m\right\rangle
\leftrightarrow \left\vert n+q,l+1,m\right\rangle $, which are coupled due
to the driving field via normalized coupling coefficients

\begin{equation}
c_{n,l,m}^{\pm q}=\frac{\left\langle n,l,m\left\vert \hat{z}\right\vert n\pm
q,l\pm 1,m\right\rangle }{C_{0}},
\end{equation}%
where $C_{0}=\left\vert \left\langle n_{0},n_{0}-1,m\left\vert \hat{z}%
\right\vert n_{0}+q,n_{0},m\right\rangle \right\vert $. Note that because of
the z-polarization of the driving field, $m$ is conserved throughout the
evolution while $l$ is only coupled to $l\pm 1$. Due to the strong
nonlinearity of the coupling coefficients and $E_{n}$ with respect to $n$,
many quantities in the problem may change by orders of magnitude when $n$
varies. Therefore, every parametrization will always be, in some sense,
local - helping one to study the vicinity of a specific value of $n$.

For the initial condition comprised of a single $\left\vert
n_{0},n_{0}-1,m\right\rangle $ CRS, one can identify three time scales in
the initial setting of the problem, i.e., the nonlinearity time scale $%
T_{nl}=q^{2}\left\vert \frac{d^{2}E_{n_{0}}}{dn_{0}^{2}}\right\vert /\hbar
\alpha $ approximating the time between the first two successive
transitions, the frequency sweep time scale $T_{s}=\alpha ^{-1/2}$ and the
Rabi time scale $T_{R}=2\hbar /C_{0}\varepsilon $. Using these three
timescales we define the dimensionless time $\tau =t/T_{s}=t\sqrt{\alpha }$
and two dimensionless parameters

\begin{equation}
P_{1}=\frac{T_{s}}{T_{R}}=\frac{C_{0}\varepsilon }{2\hbar \sqrt{\alpha }},
\label{P1}
\end{equation}
\begin{equation}
P_{2}=\frac{T_{nl}}{T_{s}}=\frac{6q^{2}R_{y}}{\hbar \sqrt{\alpha }n_{0}^{4}},
\label{P2}
\end{equation}%
characterizing the driving strength and the nonlinearity in the problem,
respectively. The parameters $P_{1,2}$ fully define the evolution of the system.
Indeed, upon expansion of the wave function in terms of the eigenfunctions $%
\left\vert \psi \right\rangle =\sum_{n,l,m}a_{n,l,m}\left\vert
n,l,m\right\rangle $ one can write the dimensionless Schrodinger equation
for the coefficients $a_{n,l,m}$

\begin{equation}
i\frac{da_{n,l,m}}{d\tau }=\overline{E}_{n}a_{n,l,m}+2P_{1}\cos {\phi _{d}}%
\sum_{n^{\prime }}\sum_{\Delta l=\pm 1}c_{n,l,m}^{n^{\prime },l^{\prime
}}a_{n^{\prime },l^{\prime },m},  \label{OrigEq}
\end{equation}%
where now $\overline{E}_{n}$ is the dimensionless energy $%
-P_{2}n_{0}^{4}/6q^{2}n^{2}$, $l^{\prime }=l+\Delta l$ and the summation
over $n^{\prime },\Delta l$ follows the restrictions on the quantum numbers,
i.e, $n^{\prime }\geq 1$, $0\leq l^{\prime }<n^{\prime }$, $\left\vert
m\right\vert \leq l^{\prime }$.

The resonant dynamics emerging from Eq. (\ref{OrigEq}) are the main
focus of this work, and it is helpful to first examine the different types
of evolutions one can expect when changing the parameters $P_{1,2}$. As a
representative example, we choose the $1:1$ resonance, with a single CRS
initial condition: $n_{0}=40$, $l_{0}=n_{0}-1=39$, $m=l_{0}=39$. Figure \ref%
{FigLC} shows the numerical solution of Eq. (\ref{OrigEq}) for $P_{1}=1$, $%
P_{2}=30$ (For details on the numerical simulations see Appendix \ref{AppB}). At
time $\tau =0$ the driving frequency passes the resonance with the Keplerian
frequency $d\overline{E}_{n}/dn$ associated with $n_{0}$.

\begin{figure}[tbp]
\includegraphics[width=3.375in]{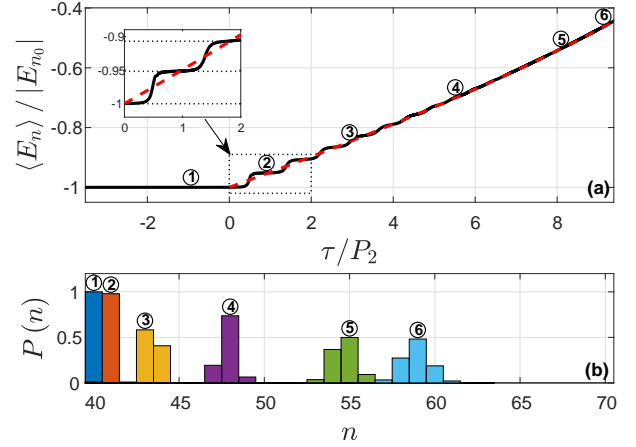}
\caption{Quantum mechanical numerical simulation of the $1:1$ resonance
dynamics for an initial CRS with $n_{0}=40$, $m=39$. The parameters are $%
P_{1}=1$,$P_{2}=30$. (a) Expectation value of the unperturbed energy as
a function of time, normalized by the magnitude of the initial energy (solid
black line). The dashed red line represents the resonance condition, (\protect
\ref{ResCond}). Inset: Zoom-in of the region in the dotted box. Dotted
lines show the normalized energy of states $%
n_{0},n_{0}+1,n_{0}+2$. (b) Population of states with different $n$'s, for
six times during the evolution as defined in (a).}
\label{FigLC}
\end{figure}
Figure \ref{FigLC}(a) shows the average unperturbed energy $\left\langle
\overline{E}_{n}\right\rangle =\sum_{n,l}\overline{E}_{n}\left\vert
a_{n,l,m}\right\vert ^{2}$ normalized with respect to the initial energy $%
\left\vert \overline{E}_{n0}\right\vert $, as a function of time and one can
see a continuing increase in the energy of the system at later times.
Furthermore, the initial growth of the energy proceeds in sharp "jumps" as
highlighted in the inset. The dashed lines in the inset show the unperturbed
energies $\overline{E}_{n_{0}},\overline{E}_{n_{0}+1},\overline{E}_{n_{0}+2}$
normalized by $\left\vert \overline{E}_{n_{0}}\right\vert $. The sharp
transitions between these energy values indicate a full population transfer
between neighboring $n$ states. As a further illustration, Fig. \ref{FigLC}(b) shows
the distribution of the population between the different $n$ values $%
P_{n}=\sum_{l}\left\vert a_{n,l,m}\right\vert ^{2}$, at six specific times,
corresponding to markers 1,2,..,6 in Fig. \ref{FigLC}(a). Comparing the distributions
at times 1 and 2 in the figure, one can observe a full population transfer
between $n_{0}$ and $n_{0}+1$ states. This trend continues at time 3 which
is in the middle of a two-level transition. However, later the energy
growth smooths, and the distribution $P\left( n\right) $ broadens -
revealing multilevel transitions in the system.

\begin{figure}[tbp]
\includegraphics[width=3.375in]{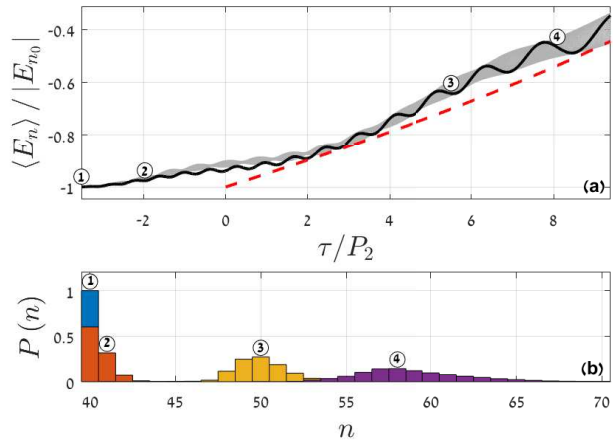}
\caption{The same as Fig. \ref{FigLC}, but for $P_{2}=0.6$. The
gray area in (a) is comprised of $100$ classical trajectories
with corresponding classical initial conditions, and uniformly distributed
initial driving phases. The dashed red line in (a) represents the resonance condition, (\protect
\ref{ResCond}).}
\label{FigAR}
\end{figure}

Next, we compare the results in Fig. \ref{FigLC} with those in Fig. \ref%
{FigAR}, corresponding to the same initial conditions and parameters, but $%
P_{2}=30$ instead of $0.6$. Figure \ref{FigAR}(a) still exhibits an energy
increase, but no sharp "jumps" associated with two-level transitions are
observed. Additionally, the distributions $P\left( n\right) $ in Fig. \ref%
{FigAR}(b) become wide shortly after the beginning of the process and much
wider than in Fig. \ref{FigAR}(b) at the final times. Such wide distributions
are indicative of classical-like behavior. This is illustrated in Fig. \ref%
{FigAR}(a), where the gray shaded area represents $100$ classical trajectories
with the corresponding initial conditions and uniformly distributed initial
driving phases (the details of the classical simulation are given in Appendix %
\ref{AppB}) and one observes that the quantum-mechanical energy of the
system with $P_{2}=30$ follows closely the classical evolution. It should be
noted that the oscillations visible in the quantum-mechanical solution in
this case have twice the driving frequency, and, thus, represent a
non-resonant effect.

In the following sections we show that indeed the dynamics in Fig. \ref%
{FigLC} represents a quantum-mechanical LC process, while that in Fig. \ref%
{FigAR} corresponds to the classical AR. We describe how Eq. (\ref%
{OrigEq}) yields the aforementioned two types of evolutions and discuss the
resonant excitation efficiency in the problem in our $P_{1,2}$ parameter
space.

%%%%%%%%%%%%%%%%  Section 2  %%%%%%%%%%%%%%%%%%%%%%%%%%%%%%%%

\section{Resonant Structure}

\label{Sec2}

We have illustrated above that the response of our system to the chirped
drive is dominated by resonant interactions. These resonances can be studied
conveniently via transformation to a rotating frame of reference, and
application of the rotating wave approximation (RWA) to neglect all rapidly
oscillating terms in Eq. (\ref{OrigEq}). To this end, we define $%
b_{n,l,m}=a_{n,l,m}e^{il\phi _{d}}$ which transforms Eq. (\ref{OrigEq}) into:

\begin{equation}
i\frac{db_{n,l}}{d\tau }\approx -\Gamma _{n,l}b_{n,l}+P_{1}\left[
c_{n,l}^{+q}b_{n+q,l+1}+c_{n,l}^{-q}b_{n-q,l-1}\right] ,  \label{RWAEq}
\end{equation}%
where index $m$ was omitted for brevity and $\Gamma _{n,l}=\overline{E}%
_{n}+l\omega _{d}$. In this rotating frame of reference only the states with
similar pseudoenergies $\Gamma _{n,l}$ are resonant, while all other states
oscillate rapidly and can be neglected. The resonance condition between
states $n,l$ and states $n+q,l+1$ is then given by equating $\Gamma _{n,l}=\Gamma
_{n+q,l+1}$. In the limit of large $n$, one finds this condition to be

\begin{equation}
\omega _{d}\approx q\frac{d\overline{E}_{n}}{dn},  \label{ResCond}
\end{equation}%
which as mentioned above, corresponds to a $q:1$ ratio between the driving
frequency and the Keplerian frequency $d\overline{E}_{n}/dn$. Since $\omega
_{d}=\omega _{0}-\tau $ (here and below we use $\omega _{d}$ and $\omega _{0}
$ normalized by $1/T_{s}=\sqrt{\alpha }$), it is possible to solve for the
value of $n$ satisfying the resonance condition, (\ref{ResCond}), as a
function of the time and use it to define the resonant value for the energy. The
dashed red lines in Figs. \ref{FigLC} and \ref{FigAR} show this resonant
energy as a function of the time and we observe that the evolution of the energy
of the system follows closely the resonant energy.

When neglecting rapidly oscillating terms in Eq. (\ref{RWAEq}), one must
verify that other pseudo-energy crossings (for different resonance ratios $q$%
) do not interfere with the desired resonant chain. To this end, we define
the time $\tau _{n}^{q}$ when the resonance condition, (\ref{ResCond}), is met
for the transition $n,l\leftrightarrow n+q,l+1$. Note that since $d\overline{%
E}_{n}/dn$ is monotonic in $n$, the resonant transitions along a given
resonant chain are ordered consecutively (i.e., $\tau _{n}^{q}<\tau
_{n+q}^{q} $). Nevertheless, if $\tau _{n+q}^{q+1}$ is larger than $\tau
_{n}^{q}$, but smaller than $\tau _{n+q}^{q}$, the two resonant chains will
mix. This leads to two conditions which must be met to avoid this mixing,
i.e., $\tau _{n+q}^{q+1}>\tau _{n}^{q}$ and $\tau _{n+q}^{q}<\tau
_{n+q}^{q+1}$. One can show that the first condition is always met, while
the second is only true starting from some minimal $n$ value. This minimal $%
n $ can be found numerically, and above this $n$ the resonant chains do not
mix. For example, the minimal $n$ is $6$, $17$ and $34$ for $q=1,2,$ and $3$,
respectively \cite{SideNote1}. Within the RWA, when the resonant chains are
separated, one can study them individually, and this is what we do next.

%%%%%%%%%%%%%%%%  Section 3  %%%%%%%%%%%%%%%%%%%%%%%%%%%%%%%%

\section{1:1 Resonance}

\label{Sec3}

\begin{figure}[t]
\includegraphics[width=3.375in]{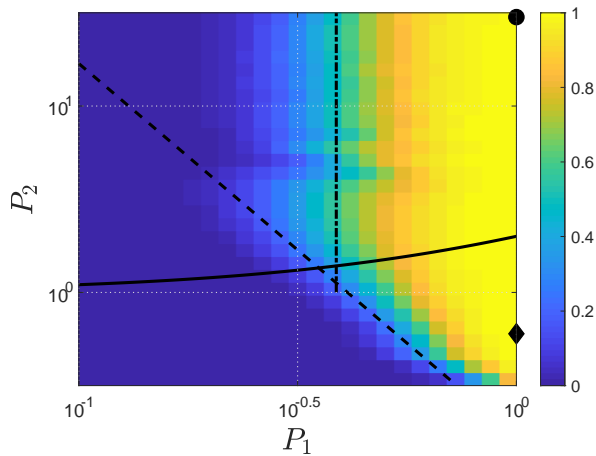}
\caption{Quantum-mechanical numerical simulations for the $1:1$ resonance case. Equation (\protect\ref{RWAEq}) is solved for the fraction of population with $n>50$ when
starting from a CRS with $n_{0}=40$ and $m=39$. The circle (diamond) show
the parameters used in Fig. \protect\ref{FigLC} (Fig. \protect\ref{FigAR}). The initial (final) driving frequency matches
the resonant frequency at $n=30$ ($n=60$). The dashed line mark the classical threshold, (\ref{CTh}). The solid line mark the quantum-classical separation line, (\ref{LCCond2}). The dashed-dotted line marks the value of $P_{1,th}^{LC}$ as calculated using Eq. (\ref{PLC}).}
\label{Fig1-1a}
\end{figure}

Section \ref{Sec1} illustrated that parameter $P_{2}$ may change the nature
of the evolution of the system, and that two efficient excitation regimes are possible. However, such efficient excitation is not guaranteed, and depends on the choice of parameters $P_{1,2}$. To further understand this effect, we 
now discuss the excitation efficiency of the $1:1$ resonance for a CRS initial condition with maximal $m$, i.e. $m=l_{0}=n_{0}-1$. As a measure for the efficiency we examine the
fraction of excited population with $n$ exceeding a certain threshold value $
n_{th}$ at the final time of evolution.
 We proceed and show the numerical solution of Eq. (\ref{RWAEq}) for the distribution of
the excitation efficiency in $P_{1,2}$ parameter space in Fig. \ref{Fig1-1a}%
(a). The initial (final) driving frequency in these simulations is given by Eq. (\ref{ResCond}) for $n=30$ ($n=60$), while $n_{th}=50$
and the initial condition is a single CRS with $n_{0}=40$ and $m=l_{0}=39$. The choice of $n_{0}$ puts the transition frequencies in the readily accessible microwave regime, while the choice of $n_{th}$ is discussed in Sec. \ref{SubSecIon}.

As expected, the higher the driving amplitude (characterized by $P_{1}$),
the higher the excitation efficiency, up to $100\%$. However, the gradual
transition from no excitation to full excitation happens in the vicinity of
two distinct lines in the parameter space represented by the dashed diagonal
line and the dashed-dotted vertical line. Clearly, this hints at two
different resonant mechanisms in play and their study is our next goal.

\subsection{Quantum Mechanical Ladder-Climbing}

\label{SubSecLC}

Motivated by the two-level transitions shown in Fig. \ref{FigLC}, we
analyze Eq. (\ref{RWAEq}) again, but this time for two neighboring levels
only,

\begin{equation}
i\frac{d}{d\tau }\left(
\begin{array}{c}
b_{n} \\
b_{n+1}%
\end{array}%
\right) =\left(
\begin{array}{cc}
\overline{E}_{n}-\left( n-1\right) \tau  & P_{1}c_{n}^{+1} \\
P_{1}c_{n}^{+1} & \overline{E}_{n+1}-n\tau
\end{array}%
\right) \left(
\begin{array}{c}
b_{n} \\
b_{n+1}%
\end{array}%
\right) ,  \label{2Level}
\end{equation}%
where index $l$ was omitted because the difference $n-l$ is conserved at $1$
along the $1:1$ resonance chain starting from a CRS. The initial driving
frequency was omitted from $\omega _{d}$ in (\ref{2Level}) for brevity, as
it could be canceled by shifting time. Equation (\ref{2Level}) describes a
two-level Landau-Zener transition \cite{Landau,Zener}. If the transitions'
times, as found from Eq. (\ref{ResCond}), are well separated the system will
undergo successive LZ transitions, commonly known as quantum energy LC. This
explains the initial dynamics shown in Fig. \ref{FigLC}(a). To see the
relevance of the LC process to the parameter space of Fig. \ref{Fig1-1a}
one needs to examine the efficiency of the process. The efficiency of a
single LZ transition, i.e., the fraction of the population transferring from
level $n$ to level $n+1$, depends on $P_{1}$ only and is given by the
LZ formula $1-\exp {\left[ -2\pi \left( P_{1}c_{n}^{+1}\right) ^{2}%
\right] }$. Indeed,\ one can see in Fig. \ref{Fig1-1a} that for large
values of $P_{2}$, the efficiency of the excitation is independent of $P_{2}$%
. Furthermore, one can find the efficiency of the full LC process, by
multiplying the efficiencies of successive single transitions:

\begin{equation}
P=\prod_{n=n_{0}}^{n_{th}}1-\exp {\left[ -2\pi \left( P_{1}c_{n}^{+1}\right)
^{2}\right] }.  \label{PLC}
\end{equation}%
By setting Eq. (\ref{PLC}) equal to $1/2$ we can define the threshold value $%
P_{1,th}^{LC}$ for which half the population will reach the target state $%
n_{th}$. In principle, this value depends on $n_{th}$, but for the
parameters of this problem within a few transitions $c_{n}^{+1}$ scales as $%
n^{2}/n_{0}^{3/2}$ , so the product (\ref{PLC}) converges rapidly and only
weakly depends on $n_{th}$. One finds numerically that $P_{1,th}^{LC}\approx
0.39$, and this value is plotted as a dashed-dotted line in Fig. \ref%
{Fig1-1a}, showing a good agreement with the numerical simulations when $%
P_{2}$ is sufficiently large.

It should be noted that our choice of initial conditions in a circular state
is not incidental. Indeed, since $m=n_{0}-1$ is conserved, the $n_{0}$ CRS
is not "connected" from below to any other state (it is the "ground state"
of the resonant chain). Therefore, the LZ transitions can only transfer the
population up the resonant chain. However, if the initial conditions were
chosen such that there existed a state below $n_{0}$, the sweeping driving
frequency would have driven the population down to this state, and the
excitation process would have stopped.

Lastly, one still needs to find the values of $P_{2}$ for which the LC
framework is applicable. As mentioned above, the LZ transitions must be well
separated in time so they can be treated as separate two-level
transitions. To check when this condition is met, we follow the footsteps of
\cite{Lit19} and \cite{Ido} and compare the time between two successive LZ transitions
and the time-width of a single transition. The width of a single LZ
transition can be estimated as $\Delta \tau _{LZ}=1+P_{1}c_{n}^{+1}$ \cite%
{Ido}, while the time between two successive transitions can be found using
Eq. (\ref{ResCond}) yielding $\Delta \tau _{between}\approx d^{2}\overline{E}%
_{n}/dn^{2}$. Therefore, the condition $\Delta \tau _{between}\gg \Delta \tau
_{LZ}$  guarantees that the transitions are well separated. Explicitly, the
condition reads:

\begin{equation}
P_{2}\left( \frac{n_{0}}{n}\right) ^{4}\gg 1+P_{1}c_{n}^{+1},  \label{LCCond}
\end{equation}%
where again index $l$ was omitted. For several initial LZ steps starting
from some $n_{0}\gg 1$, this condition can be relaxed by substituting $n_{0}$
for $n$ and recalling that by construction $c_{n_{0}}^{+1}=1:$

\begin{equation}
P_{2}\gg 1+P_{1}.  \label{LCCond2}
\end{equation}%
The solid line in the parameter space in Fig. \ref{Fig1-1a} represents $%
P_{2}=1+P_{1}$ and one can see that above this line the efficiency depends
only on $P_{1}$, while $P_{1,th}^{LC}$ (dashed-dotted line in Fig. \ref{Fig1-1a})
bounds the region of an efficient LC process. Returning to Fig. \ref{FigLC}%
 which exhibits LC type evolution, one can see that its $P_{1,2}$
parameters are well inside the quantum region (the parameters are marked by
a black circle in Fig. \ref{Fig1-1a}). If condition (\ref{LCCond2}) is
not satisfied, the transitions are not well separated and many states mix.
This type of evolution is demonstrated in Fig. \ref{FigAR}, where the
parameters are well below the separation line (marked by the black diamond  in
Fig. \ref{Fig1-1a}). The nature of the evolution in this case is studied
next.

\subsection{Classical Autoresonance}

\label{SubSecAR}

The previous subsection showed that quantum-mechanical analysis does not fully explain the results in Fig. \ref{Fig1-1a}. One can see in the figure two different threshold lines for efficient excitation. We show now that the dashed diagonal line corresponds to classical evolution. To understand this region in the parameter space, we turn to the classical analysis in Ref. \cite{Gros}. We present the main results of \cite{Gros} here for
completeness, while reformulating this theory in terms of our dimensionless quantum-mechanical parameters $P_{1,2}$. 

\begin{figure*}[t]
\includegraphics[width=6.75in]{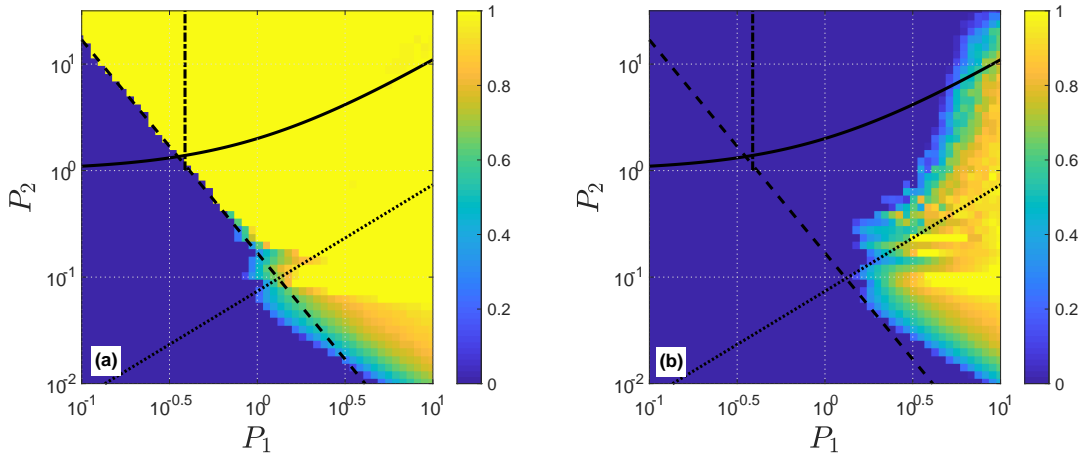}
\caption{Classical numerical simulations for the $1:1$ resonance case, with parameters and initial conditions corresponding to Fig. \ref{Fig1-1a}. (a)
Solution of the classical equations of motion for the fraction of initial
conditions leading to $\bar{I}_{3}>50$ when
starting from a circular orbit with $\bar{I}_{3,0}=40$ and $\bar{I}_{1,0}=40$%
. (b) Same as (a) but for the fraction of initial conditions leading to
ionization. The initial (final) driving frequency matches
the resonant frequency at $n=30$ ($n=60$). Dashed lines mark the classical threshold (\ref{CTh}). Solid lines mark the quantum-classical separation line (\ref{LCCond2}). Dotted lines mark the breakdown of the SRA, (\ref{SRABreak}), for $\gamma=1$,$n=n_{f}$. Dashed-dotted lines mark the value of $P_{1,th}^{LC}$ as calculated using Eq. (\ref{PLC}).}
\label{Fig1-1b}
\end{figure*}

The classical problem of the driven atom is conveniently
analyzed using the three pairs of action-angle variables of the unperturbed
problem. The actions $I_{2},I_{1}$ are associated with the total angular
momentum and its projection on the $z$ axis, respectively, while the action $%
I_{3}$ characterizes the unperturbed Hamiltonian which is proportional to $%
I_{3}^{-2}$. The semi-classical approximation then yields $I_{3}\approx
\hbar n$, $I_{2}\approx \hbar l$, $I_{1}\approx \hbar m$.
The classical theory in \cite{Gros} used the single resonance
approximation (a classical analog of the RWA). It resulted in a dimensionless Hamiltonian, which in terms of our parameters $P_{1,2}$ can be written as

\begin{equation}
H\left( \Theta _{1,2,3},\bar{I}_{1,2,3}\right) =-\frac{P_{2}n_{0}^{4}}{6\bar{%
I}_{3}^{2}}+\frac{\sqrt{2}P_{1}}{n_{0}^{3/2}}\bar{I}_{3}^{2}\sin {i}\sin {%
\Phi },  \label{ClassH}
\end{equation}%
where the actions are normalized by $\hbar $ ($\bar{I}_{1,2,3}=I_{1,2,3}/%
\hbar $), time is normalized by $1/\sqrt{\alpha }$, distances are
normalized with respect to Bohr's radius $a_{0}$, $i$ is the inclination angle ($\sin i=I_{1}/I_{2}$) and $\Phi =\Theta
_{3}+\Theta _{2}-\phi _{d}$ with $\Theta _{2,3}$ being the angle variables
corresponding to actions $I_{2,3}$. The physical meaning of the phase
mismatch $\Phi $ is revealed when examining its temporal derivative

\begin{equation}
\frac{d\Phi }{d\tau }=\frac{d\Theta _{3}}{d\tau }+\frac{d\Theta _{2}}{d\tau }%
-\omega _{d}\approx \frac{d\Theta _{3}}{d\tau }-\omega _{d}+O\left(
P_{1}\right) .
\end{equation}%
If $d\Phi /d\tau \approx 0,$ the orbital frequency $d\Theta _{3}/d\tau $
approximately follows the driving frequency. This classical resonance
condition is actually the same as the quantum one, (\ref{ResCond}), within the
semiclassical approximation. It is shown in Ref. \cite{Gros} that if the
driving frequency starts sufficiently far from the resonance, Hamiltonian (%
\ref{ClassH}) yields a continued phase-locking $\Phi \approx 0$ after
passage through resonance provided that (using our parametrization)
\begin{equation}
\sqrt{P_{2}}P_{1}>0.41.  \label{CTh}
\end{equation}%
If this sharp threshold condition is satisfied, the resulting phase locking
yields a continuous increase in the energy as the system self-adjusts to
stay in resonance for an extended period of time. Note that the form of the
left hand side in the classical threshold condition, (\ref{CTh}), could have
been predicted even without the detailed analysis in \cite{Gros}, as it is
the only combination of parameters $P_{1,2}$, which does not depend on $\hbar $
(which cancels out after we replace $n_{0}$ with the initial dimensionless $%
I_{3}$). We illustrate the sharp threshold phenomenon of the classical
autoresonance in Fig. \ref{Fig1-1b}(a), showing the excitation efficiency as a
function of parameters $P_{1,2}$ using the \textit{exact} classical
equations of motion (for details on these simulations see Appendix \ref{AppB}).
In order to check the independence of the capture of the system in
autoresonance on the initial phase mismatch, we started the simulation on a
circular orbit with spherical angles $\varphi =\theta =0$, and averaged over
the initial driving phase between $0$ and $2\pi $. All other parameters are
the same as in Fig. \ref{Fig1-1a}. One can see that both the classical [Fig. \ref{Fig1-1b}(a)] and the quantum
(Fig. \ref{Fig1-1a}) simulations correctly recreate the threshold condition (dashed line) even
though the threshold region is much narrower in the classical results.
Naturally, the classical simulations entirely ignore the quantum-classical
separation (solid line) given by condition (\ref{LCCond2}), further
demonstrating the quantum nature of the evolution above the separation line,
and identifying the dashed line as the classical threshold. The broadening
of the threshold region in Fig. \ref{Fig1-1a} can be attributed to quantum
fluctuations of the initial state, which were absent in the classical
simulations.

The previous and current subsections describe \textit{purely} LC or AR
evolutions. However, as mentioned above, one must also consider an
intermediate case where the initial evolution is of an LC nature (i.e.,
condition [\ref{LCCond2}] is met), but in the final evolution stage
condition (\ref{LCCond}) is violated and we expect a dynamical transition
from LC to AR at later times. In fact, this situation was relevant to
practically all of the region in Fig. \ref{Fig1-1a} above the
quantum-classical separation line. Nevertheless, as demonstrated in Fig. %
\ref{FigLC}(b), the efficiency of the excitation remains high and smooth
despite the transition from LC to AR. The reason for this smooth transition
can be explained by observing that the LC\ process closely follows the
resonance with the drive. In turn, this also means that at the transition to
the classical regime, the evolution is phase-locked to the drive and the
classical phase mismatch $\Phi $ remains bounded. This guarantees smooth
transition to the AR regime as the classical dynamics emerges during the
chirped excitation process.

Finally, one can see in the lower right part in Fig. \ref{Fig1-1b}(a) that
the transition region to efficient excitation is not as narrow. We attribute
this effect to the breaking of the SRA when parameter $P_{1}$ becomes large,
as discussed in the next section.

\subsection{The breaking of the single resonance approximation and ionization}

\label{SubSecIon}

Our quantum-mechanical model does not include ionization channels, so we
discuss the problem of ionization within the classical theory. Classically,
the ionization in the driven-chirped problem can occur when the SRA loses
its validity. This effect was studied in \cite{Gros}, where it is shown that
the breakdown of the SRA happens when the frequency of oscillations of $\Phi
$ in autoresonance become of the order of the driving frequency and other
resonant terms become important. When this happens the dynamics is not
dominated by the $1:1$ resonance, and ionization may soon follow via chaotic
dynamics. Based on \cite{Gros} the condition for breakdown of the SRA is

\begin{equation}
\frac{P_{1}}{P_{2}}>\gamma \frac{n_{0}^{7/2}}{9\sqrt{2}\sqrt{1-\left( \frac{m%
}{n}\right) ^{2}}n^{4}},  \label{SRABreak}
\end{equation}%
where we have used the semi-classical approximation for the dimensionless actions
and $\gamma $ is a numerical factor smaller than $1$. Condition (\ref%
{SRABreak}) is again local, and gets easier to satisfy for higher $n$.
Therefore, for estimation, we substitute $n=n_{f}$ \ in (\ref{SRABreak}),
where $n_{f}$ is the resonant value of $n$ at the end of the excitation
process. Figure \ref{Fig1-1b}(b) shows the classical ionization probability
for the same parameters as in Fig. \ref{Fig1-1b}(a). The diagonal dotted line
in the figure is given by Eq. (\ref{SRABreak}) for $n=n_{f}$ and $\gamma =1$%
. One can see that the ionization regime is centered around this line.
Furthermore, our quantum-mechanical simulations in Fig. \ref{Fig1-1a} are
performed in the portion of the parameter space for which no ionization
happens classically. In this part of the parameter space we do not expect
ionization to occur.

In order to avoid destruction of AR at large $n$ [see condition (\ref{SRABreak})], we limit the value of $n_{f}$, so that a sizable part of the parameter space avoids ionization. The value of $n_{th}$ was taken halfway between $n_{0}$ and $n_{f}$ to identify significant excitation.

%%%%%%%%%%%%%%%%  Section 3  %%%%%%%%%%%%%%%%%%%%%%%%%%%%%%%%

\section{2:1 Resonance}

\label{Sec4}

\begin{figure*}[t]
\includegraphics[width=6.75in]{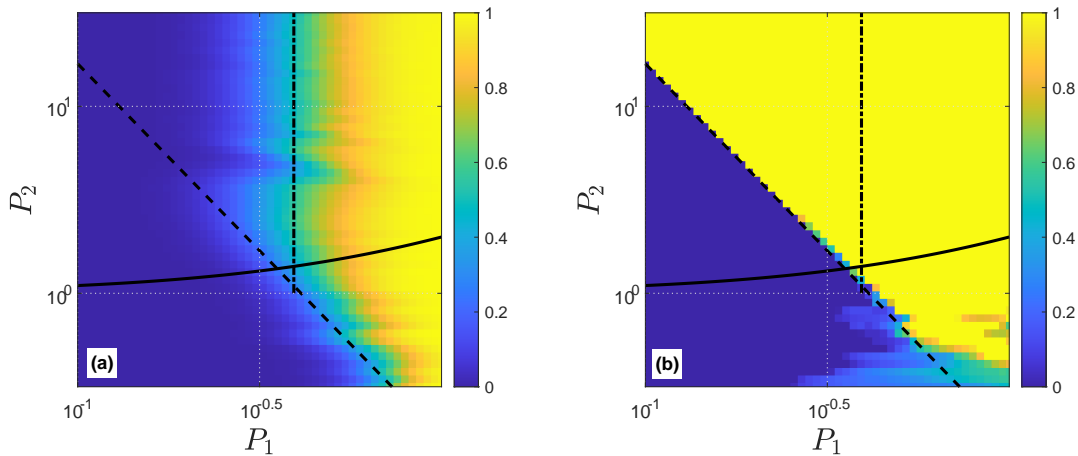}
\caption{Numerical simulations of the $2:1$ resonance. (a) Solution of Eq. (%
\protect\ref{RWAEq}) for the fraction of population with $n>100$ when
starting from a circular state with $n_{0}=90$ and $m=0$. (b) Solution of
the classical equations of motion for the fraction of initial conditions
leading to $\bar{I}_{3}>100$ when starting from a circular orbit with $\bar{I%
}_{3,0}=90$ and $\bar{I}_{1,0}=0$. In all simulations the initial (final)
driving frequency matches the resonant frequency at $n=80$ ($n=110$). Dashed lines mark the classical threshold, (\ref{CTh2}). Solid lines show the quantum-classical separation line, (\ref{LCCond2-1}) for $n=n_{0}$. Dashed-dotted lines mark the value of $P_{1,th}^{LC}$ as calculated using a modified version of Eq. (\ref{PLC}) for the $2:1$ resonance (see the text).}
\label{Fig2-1}
\end{figure*}

Section \ref{Sec3} revolved around analytic and numerical results for the $%
1:1$ resonance, but the analysis is not limited to this choice. In this
section, we show that the same considerations could be applied to the $%
2:1$ resonance, leading to similar results. Consider a CRS initial condition
defined by $n_{0}$ and $m$, such that $m$ is not restricted and can take any value $\left\vert m\right\vert <n_{0}$. The choice of $n_{0}$ and $n_{th}$ follows the same considerations as in Sec. \ref{Sec3}. The
driving frequency now sweeps through twice the Keplerian frequency i.e., $2d\overline{E%
}_{n_{0}}/dn_{0}$, and the resonant transitions are $n,l,m\leftrightarrow
n+2,l+1,m$. The analysis again starts with Eq. (\ref{RWAEq}), but now for $%
q=2$, so a two-level description similar to Eq. (\ref{2Level}) follows
immediately. The width of a single LZ transition and the time between two
transitions are found similarly to Sec. (\ref{Sec3}) and the
quantum-classical separation criterion is found to be

\begin{equation}
P_{2}\left( \frac{n_{0}}{n}\right) ^{4}\gg 1+P_{1}c_{n,l}^{+2}.
\label{LCCond2-1}
\end{equation}%
As with the $1:1$ resonance, the initial stages of the evolution are the
most important and condition (\ref{LCCond2-1}) could be replaced by its
version for $n=n_{0}$, yielding the same result as Eq. (\ref{LCCond2}).
Figure \ref{Fig2-1} shows numerical simulations for the efficiency of
excitation by passage through the $2:1$ resonance. Figure \ref{Fig2-1}(a) shows
quantum mechanical simulations for $n_{0}=90$,$m=0$, while Fig. \ref{Fig2-1}(b)
shows classical simulations for the corresponding initial condition with $%
\bar{I}_{3}=\bar{I}_{2}=90$ and $\bar{I}_{1}=0$. In the quantum simulations
the efficiency is determined by the fraction of population exceeding $n_{th}=100$%
, while for the classical simulations it is defined by the fraction of
initial conditions out of a uniformly distributed initial phases that reach
a final unperturbed energy corresponding to $\bar{I}_{3}>100$. Note that the
range of $P_{1,2}$ in Fig. \ref{Fig2-1} is the same as that in Fig. \ref%
{Fig1-1a}.

The solid lines in Figs. \ref{Fig2-1}(a) and \ref{Fig2-1}(b) separate the quantum and
classical regions of the evolution. One can again observe the two regimes in Fig. \ref{Fig2-1}(a) separated by this line, and the absence of
this separation in the classical simulation in Fig. \ref{Fig2-1}(b). The
efficiency of the LC process above this separation line could be calculated
similarly to Eq. (\ref{PLC}) by successively multiplying the efficiencies of
individual LZ transitions. Once again, because the coupling coefficients
grow rapidly, the threshold value $P_{1,th}^{LC}$ for which the total
efficiency is $0.5$ depends only weakly on the number of transitions. It
also depends rather weakly on the value of $m$. For example, the parameters $%
n_{0}=90$, $m=0$ as in Fig. \ref{Fig2-1}, yield $P_{1,th}^{LC}\approx 0.39$
(same as in Sec. \ref{Sec3}), but when $m=89$, $P_{1,th}^{LC}\approx 0.34$.
The value of $P_{1,th}^{LC}$ is represented in Figs. \ref{Fig2-1}(a) and \ref{Fig2-1}(b) by
dashed-dotted vertical lines. One can observe good agreement between the
predictions of the quantum mechanical simulation and this line in Fig. \ref%
{Fig2-1}(a).

It should be noted that like the $1:1$ resonance in Sec. \ref{Sec3}, the CRS
studied here has the property that it is not "connected" from below to any
other state along the resonant chain. The state below the initial condition
in the chain would have had $n=n_{0}-2=l$ which is not a physical state.
Since this is true for every $m$, the $2:1$ LC continuing excitation process
could be applied to any $m$, unlike the $1:1$ LC.

When condition (\ref{LCCond2}) is violated, the classical dynamics emerges
and one can use the results in \cite{Gros} to find that the capture into
classical AR is only possible when

\begin{equation}
\sqrt{P_{2}}P_{1}>0.41.  \label{CTh2}
\end{equation}%
Remarkably, this result is identical to the one observed for the $1:1$ resonance [Eq. (\ref{CTh})], demonstrating the uniqueness of the parametrization used in this work.
The dashed lines in Fig. \ref{Fig2-1} show the threshold $\sqrt{P_{2}}%
P_{1}=0.41$ for efficient excitation. The classical simulations exhibit a
sharp transition at this line, except for low $P_{2}$ as one gets closer to
the breaking of the SRA (the corresponding breaking line, as in Fig. \ref%
{Fig1-1b} is outside the range of the $P_{1,2}$ values in Fig. \ref{Fig2-1}%
). In the quantum mechanical simulations, the classical threshold is
retrieved below the quantum-classical separation line, (\ref{LCCond2}), but
is broadened compared to the classical simulations due to quantum
fluctuations.

%%%%%%%%%%%%%%%%  SUMMARY  %%%%%%%%%%%%%%%%%%%%%%%%%%%%%%%%

\section{SUMMARY}

\label{summary}

In conclusion, we have studied the problem of resonant excitation of a
Rydberg atom starting in a CRS using chirped drive. Based on three
characteristic timescales in the problem, we introduce two dimensionless
parameters, $P_{1,2}$ [Eqs. (\ref{P1}) and (\ref{P2})], and study the resonant
nature of the problem in this parameter space within the rotating-wave
approximation. We have shown how this approximation allows one to reduce the
three dimensional problem to one dimensional resonant interactions
characterized by the $q:1$ ratio ($q=1,2$) between the driving and the Keplerian
frequencies. The $1:1$ and $2:1$ resonances are studied in detail each
showing two distinct persistent resonance regimes, i.e., the
quantum-mechanical ladder climbing and the classical autoresonance. The
major criteria (borderlines) in the $P_{1,2}$ parameter space are
discussed, including (a) the separation line between the two regimes and (b)
the regions of efficient excitation in the two regimes. In both regimes very
high efficiencies ($\sim 100\%$) are possible, but the LC process yields
significantly narrower (in $n$) excited wave packets. Our analytic results
are supported by classical and quantum-mechanical numerical simulations
demonstrating the validity of our theoretical approach, as well as the
quantum-classical correspondence, and other effects such as quantum
fluctuations. The ionization process in the chirped-driven excitation is
discussed classically in the framework of breaking of the single resonance
approximation in the problem. It is shown that the ionization effect is
negligible in the areas of interest in our quantum-mechanical simulations.

The results of this work extend previous studies of the chirped-driven
Rydberg atom into the boundary between the quantum and the classical evolution.
From a broader perspective, it is also the first use of the formalism for
studying such quantum-classical transitions in a three-dimensional problem.
The processes described in this work enlarge the tool-box for the control
and manipulation of Rydberg atoms, and may lead to new applications. It will
be interesting to study this problem for other initial conditions which are
not CRS in the future. Generally speaking, such initial conditions should
not exhibit sharp thresholds for capture into AR, but, rather, a different
capture process which could be conveniently studied in phase-space \cite%
{Lit14a,phasespace}. Quantum-mechanically, such initial conditions will not
be the "ground state" of their resonant chain, so the climb up the energy
ladder would require starting close to the resonance rather than sweeping
through it. Another avenue for research could be studying time varying chirp
rates. The time between LZ transitions decreases by orders of magnitude as one
climbs up the energy ladder and, thus, lowering chirps in time may allow us to
prolong the LC process and reduce the possibility of ionization.

\begin{acknowledgments}
This work was supported by Israel Science Foundation Grant No. 30/14.
\end{acknowledgments}

%%%%%%%%%%%%%%%%%%%%%%%%%%%%%%%%%%%%%%%%%%%%%%%%%%%%%

\appendix

\section{Coupling Coefficients}

\label{AppA}

In computing the coupling coefficients $\left\langle n,l,m\left\vert \hat{z}%
\right\vert n^{\prime },l^{\prime },m^{\prime }\right\rangle $ we use the
spherical coordinates $r,\theta ,\varphi $ and separate the integral for
the coefficients into the radial and angular parts. The angular part is
found by expressing $z$ as a function of $r$ and the spherical harmonic $%
Y_{1}^{0}\left( \theta ,\varphi \right) $. The functions $\psi _{n^{\prime
},l^{\prime },m^{\prime }}$ and $\psi _{n,l,m}^{\ast }$ contribute two more
spherical harmonics, and the product of the three could be integrated in
terms of the Wigner 3j symbol, yielding the angular contribution, as well as
the selection rules $m=m^{\prime }$ and $l=l^{\prime }\pm 1$. For the radial
part we first normalize $r$ by $m_{e}a_{0}/2\mu $, where $a_{0}$ is Bohr's
radius and $\mu $ the reduced mass (the normalization factor is absorbed
into $\varepsilon $). The radial integral is then given by

\begin{gather*}
\int_{0}^{\infty }r^{3}R_{n,l}^{\ast }\left( r\right) R_{n^{\prime
},l^{\prime }}\left( r\right) dr, \\
R_{n,l}\left( r\right) =\sqrt{\frac{\left( n-l-1\right) !}{2n^{4}\left[
\left( n+l\right) !\right] }}e^{-\frac{r}{2n}}\left( \frac{r}{n}\right)
^{l}L_{n-l-1}^{2l+1}\left( \frac{r}{n}\right) ,
\end{gather*}%
where $L_{a}^{b}$ is the generalized Laguerre polynomial. Note that $%
L_{a}^{b}$ is a polynomial of order $a$, and therefore the product $%
r^{3}R_{n,l}^{\ast }\left( r\right) R_{n^{\prime },l^{\prime }}\left(
r\right) $ could be broken into the sum of $(n-l-1)\times (n^{\prime
}-l^{\prime }-1)$ terms proportional to $r^{k}e^{-rp}$, where $k,p>0$. The
integral for each term yields $p^{-1-k}\Gamma \left( 1+k\right) $, with $%
\Gamma $ the Euler Gamma function. The final result reads

\begin{multline*}
c_{n,l,m}^{n^{\prime },l+1}=\frac{1}{2}\sqrt{\frac{\left( l-m+1\right)
\left( l+m+1\right) }{\left( 2l+3\right) \left( 2l+5\right) }}\times \\
\times \sqrt{\left( n-l-1\right) !\left( n^{\prime }-l-2\right) !\left(
n+l\right) !\left( n^{\prime }+l+1\right) !}\times \\
\times \sum_{i=0}^{n-l-1}\sum_{j=0}^{n^{\prime
}-l-2}f_{i}^{n,l}f_{j}^{n^{\prime },l+1}D,
\end{multline*}%
where
\begin{equation*}
\begin{array}{ccc}
D & = & \left( \frac{2nn^{\prime }}{n+n^{\prime }}\right) ^{2l+5+i+j}\left(
2l+4+i+j\right) !, \\
f_{i}^{n,l} & = & \left( -1\right) ^{i}\left[ n^{i+l+2}\left( n-l-1-i\right)
!\left( 2l+1+i\right) !\left( i\right) !\right] ^{-1}.%
\end{array}%
\end{equation*}%
These $c_{n,l,m}^{n^{\prime },l+1}$ were computed using a symbolic software,
to avoid numerical accuracy issues. Note that the coupling of CRSs to other
CRSs, or nearly circular states, contains only a small number of
contributions and can by calculated explicitly. Namely, in the limit $%
n_{0}\gg 1$ the value of $C_{0}$ is $\sqrt{2}n_{0}^{3/2}$ for the $1:1$
resonance with $m=n_{0}-1$, and $\sqrt{1-\left( m/n_{0}\right) ^{2}}%
n_{0}^{3/2}/\sqrt{2}$ for the $2:1$ resonance.

%%%%%%%%%%%%%%%%%%%%%%%%%%%%%%%%%%%%%%%%%%%%%%%%%%%%%

\section{Numerical Simulations}

\label{AppB}

The quantum mechanical simulations in Figs. \ref{FigLC} and \ref{FigAR} use
Eq. (\ref{OrigEq}). The maximal value of $n$ and $n-l$ was chosen such that
only a negligible portion of the population reaches the states along those
numerical boundaries in the Hilbert space. The simulations in Figs. \ref%
{Fig1-1a} and \ref{Fig2-1}(a), however, are based on the RWA [Eq. (\ref%
{RWAEq})], i.e., include only the states which are connected to the initial
condition through the resonant interaction. This validity of this assumption
improves as $P_{2}$ increases and breaks down completely in the portion of
the parameter space where ionization occurs. For this reason the quantum
mechanical simulations are limited to the region of the parameter space
where no ionization is observed (classically). For the $1:1$ resonance we
have also tested the effect of the RWA by solving the same equation set with
more states outside the resonant chain (i.e., states with higher values of $n-l
$) and found no significant changes in the results presented in Fig. \ref%
{Fig1-1a}.

Our classical simulations are based on solving the classical Hamilton
equations for the Hamiltonian:

\begin{equation*}
H=\frac{P_{2}n_{0}^{4}}{6q^{2}}\left[ p_{r}^{2}+\frac{p_{\theta }^{2}}{r^{2}}%
+\frac{p_{\phi }^{2}}{r^{2}\sin ^{2}\theta }-\frac{2}{r}\right] +\frac{2P_{1}%
}{C_{0}}\cos {\phi _{d}}r\cos \theta ,
\end{equation*}%
where $r,\theta ,\phi $ are spherical coordinates and $p_{r},p_{\theta
},p_{\phi }$ their conjugate momenta. Naturally, the quantum mechanical
initial condition does not translate directly to a classical initial
condition. We used initial conditions corresponding to a classical circular
Keplerian case, but averaged over the initial driving phase $\phi _{d}$
between $0$ and $2\pi $ in Fig. \ref{Fig1-1b} and over $\theta $ between
$0$ and $\pi $ in Fig. \ref{Fig2-1}(b) for testing the validity of the
single resonance approximation. One can observe that in both figures all
initial conditions yield the same results except for the bottom-right corner
of the parameter space where the SRA is not valid.

%%%%%%%%%%%%%%%%   bib   %%%%%%%%%%%%%%%%%%%%%%%%%%%%%%%%


\begin{thebibliography}{99}
\bibitem{SideNote1} To calculate the exact values of the minimal $n$ value,
one has to use the full form of $\tau_{n}^{q}$, which fulfills exactly $%
\Gamma_{n,l}=\Gamma_{n+q,l+1}$, and not the approximated form derived from
Eq. (\ref{ResCond}).
\end{thebibliography}
\end{document}